\documentclass[3p,times]{elsarticle}
\usepackage{hyperref}
\usepackage[english]{babel}
\usepackage{graphicx}
\usepackage{caption}
\usepackage{subcaption}

\usepackage[utf8]{inputenc}
\usepackage{amsmath}
\usepackage{graphicx}
\usepackage[colorinlistoftodos]{todonotes}
\usepackage{longtable}

\newcommand{\nnb}{\nonumber}

\newcommand{\ice}[1]{\relax}
\newcommand{\vep}{\varepsilon}
\newcommand{\vphi}{\varphi}
\newcommand{\Ga}{\Gamma}

\newcommand{\wdt}{\widetilde}
\newcommand{\beq}{\begin{equation}}
\newcommand{\eeq}{\end{equation}}
\newcommand{\bea}{\begin{eqnarray}}
\newcommand{\eea}{\end{eqnarray}}

\newcommand{\al}{\alpha}
\newcommand{\be}{\beta}
\newcommand{\ga}{\gamma}

\newcommand{\ep}{\varepsilon}

\ice{
\newcommand{\Fgamma}{F_\gamma}
\newcommand{\Fgammap}{F_{\gamma'}}
\newcommand{\Fgammaq}{F_\gamma^q}
}

\newcommand{\Fgamma}{\langle \gamma \rangle}
\newcommand{\Fgammap}{\langle \gamma' \rangle}
\newcommand{\Fgammaq}{\langle \gamma^q \rangle}

\newcommand{\FGamma}{\langle \Gamma \rangle}
\newcommand{\FGammaq}{\langle \Gamma^q \rangle}
\newcommand{\FGammap}{\langle \Gamma^p \rangle}

\newcommand{\ccu}{g^}

\usepackage{lineno}

\begin{document}

\title{Six loop analytical calculation of the field  anomalous dimension and
  the critical exponent $\eta$ in $O(n)$-symmetric $\vphi^4$ model}

\author[spbu]{D.V.~Batkovich}
\ead{batya239@gmail.com}
\author[kit]{K.G.~Chetyrkin}
\ead{konstantin.chetyrkin@kit.edu}
\author[spbu]{M.V.~Kompaniets\corref{cor1}}
\ead{m.kompaniets@spbu.ru}

\address[spbu]{ St. Petersburg State University,
7/9 Universitetskaya nab., 
St. Petersburg, 199034 
Russia.}
\ice{
\address[kit]{Karlsruhe Institute of Technology}
}
\address[kit]{Institut f\"ur Theoretische Teilchenphysik, Karlsruher
  Institut f\"ur  Technologie (KIT), D-76128 Karlsruhe, Germany}

\cortext[cor1]{Corresponding author}


\date{\today}

\begin{abstract}

We report on a completely analytical calculation of the field
anomalous dimension $\gamma_{\vphi}$ and the critical exponent $\eta$
for the $O(n)$-symmetric $\vphi^4$ model at the record six loop level.
We successfully compare our result for $\gamma_{\vphi}$ with $n=1$ with
the predictions based on the method of the Borel resummation combined
with a conformal mapping [Kazakov/Shirkov/Tarasov (1979)].
Predictions for seven loop
contribution to the field anomalous dimensions are given.
\end{abstract}

\maketitle

\section{Introduction}

Since Kenneth Wilson, who was first to apply $\epsilon$-expansion and
renormalization group method to calculate critical exponents in $\vphi^4$
model,
this model became one of the most popular testing grounds for a wide range of
methods of diagram calculations and resummation. The first two terms of the
$\epsilon$-expansion were calculated by Wilson in \cite{Wilson72},
$\epsilon^3$ terms and $\epsilon^4$ for critical exponent $\eta$ were 
calculated in \cite{BGZN73}.  The latter work was the last where calculations
using Wilson renormalization group approach were performed for this
model. All subsequent calculations were performed using quantum field
renormalization group approach, which  effectively reduces the  problem of
evaluation of  critical exponents to the one of finding  the  corresponding 
beta-function (or the  anomalous dimension). 

This approach combined with modern computational techniques allows one to
calculate high order corrections with significantly less effort than in the original
Wilson's formalism.  Using this approach $\epsilon^4$ terms for other
exponents were found in \cite{VKT79}.  The field anomalous dimension
$\gamma_{\vphi}$ and the critical exponent $\eta$  were  calculated
with 5-loop accuracy in \cite{phi412}, the 5-loop $\beta$-function was first
published in \cite{phi434,uniquenessK}. Later some (numerically
insignificant) inaccuracies were found in this calculation  and results for
index $\eta$ and $\beta$-function were corrected \cite{phi456}. Recently,
a completely independent check of the analytic results
\cite{phi412,phi434,uniquenessK,phi456} was successfully performed in
\cite{phi4n} with the use of purely numerical methods.

In this work we describe the results of a completely analytical calculation
of $\gamma_{\vphi}$ and  $\eta$ at six loop level in the $O(n)$-symmetric
$\varphi^4$ model.

\section{ Setup and notations}

The (renormalized) Lagrangian of the  $\varphi^4$-model in the  Euclidean space
of $d=4-2\varepsilon$ dimensions reads
\begin{equation}
{\cal L} (\varphi) = 
\frac{1}{2}m^2 Z_1\varphi^2 
+
\frac{1}{2}Z_2\left(\partial\varphi \right)^2
+\frac{16\, \pi^2}{4!}Z_4 \, g\, \mu^{2\varepsilon}\, \varphi^4,
\label{Sr}
\end{equation}
where  RCs  (Renormalization Constants)  $Z_i$ are expressed in terms of
renormalization constants of the field $\varphi_0=\varphi Z_\varphi$, mass
$m_0^2=m^2 Z_{m^2}$ and coupling constant $g_0 = g\mu^{2\epsilon} Z_g$ in the
standard way:
\begin{equation}\label{ZZ}
Z_1=Z_{m^2}Z_{\varphi}^2,\qquad Z_2=Z_{\varphi}^2, \qquad Z_4=Z_{g} Z_{\varphi}^4.
\end{equation}

\ice{ Term proportional to $\delta m^2$ is usually interpreted as a shift of
  the critical temperature and originates from $\Lambda$-renormalization. This
  term can be omitted while performing calculations in ``formal scheme''
  \cite{Vasiliev}.  } 


\ice{
In the MS-scheme \cite{'tHooft:1973mm} which we employ throughout the paper
the UV counterterms and, consequently, all renormalization constants \todo{
  one can understand this phrase as RCs depend polynomially on $p$ and $m$ }
do not depend on $\mu$ and may depend only {\em polynomially} on any other
dimensionfull parameter of a theory \cite{Collins:1974da}. As result the RCs
$Z_i$ do depend on the regulating parameter $\varepsilon$ {\em only} and can
be written as: 
}
In the MS-scheme \cite{'tHooft:1973mm} which we
  employ throughout the paper the UV counterterms do not depend on $\mu$ and
  may depend only {\em polynomially} on any other dimensionfull parameter of a
  theory \cite{Collins:1974da}. As a result the RCs $Z_i$ do depend on the
  regulating parameter $\varepsilon$ and renormalized coupling constant $g$
  {\em only} and can be written as:
\beq
Z_i = 1+ \sum_{k=1} \frac{Z_{i,k}(g)}{\vep^k}
\eeq
Given the  RC $Z_{\vphi}(g)$, the corresponding anomalous dimension of the scalar field we are
interested in  is defined as follows
\ice{
\beq
\gamma_{\vphi}(g) = \mu^2\frac{\mathrm{d}\, \log Z_{\vphi}(g)}{\mathrm{d}\, \mu^2}|{{}_{g_0,\vphi_0}}
\equiv  g \frac{\mathrm{d} \, Z_{\vphi,1}(g)}{\mathrm{d} \,g} \equiv  \frac{1}{2} g \, \frac{\mathrm{d}\, Z_{2,1}(g)}{\mathrm{d} g}
{}.
\label{g_phi}
\eeq  
}
\beq
\gamma_{\vphi}(g) = \mu\frac{\partial\, \log Z_{\vphi}(g)}{\partial\, \mu}\Big|_{g_0,\vphi_0}
= \beta(g) \frac{\partial \log Z_\vphi}{\partial g} = -2g\frac{\partial \, Z_{\vphi,1}(g)}{\partial \,g} =  -g \, \frac{\partial\, Z_{2,1}(g)}{\partial g}
{}.
\label{g_phi}
\eeq

The RC $Z_2$ and $Z_{m^2}$ are related with UV divergences of the two point one particle irreducible Green function $\Gamma_2(p,m_0^2,g_0)$, which is connected with two point Green function (propagator) $D(p,m_0^2,g_0)$ by Dyson equation $D^{-1}(p,m_0^2,g_0)=p^2 +m_0^2-\Gamma_2(p,m_0^2,g_0)$. Thus for renormalized two point Green function $D^R(p,m^2,g,\mu)$ we got
\begin{equation}\begin{split}
D^R(p,m^2,g,\mu)=\frac{1}{Z_{\vphi}^2}D(p, m^2Z_{m^2},g\mu^{2\varepsilon}Z_g)&=\frac{1}{Z_{\vphi}^2(p^2+m^2Z_{m^2}-\Gamma_2(p,m^2Z_{m^2},g\mu^{2\varepsilon}Z_g))}=\\&=\frac{1}{p^2 Z_2 +m^2 Z_1 - Z_\vphi^2\Gamma_2(p,m^2Z_{m^2},g\mu^{2\varepsilon}Z_g)}
\end{split}
\label{DR}
\end{equation}
Last term in \eqref{DR}  can be rewritten with use of the Bogoliubov-Parasiuk R-operation \cite{bog1957,bogshirk} in the following way $Z_\vphi^2\Gamma_2(p,m^2Z_{m^2},g\mu^{2\varepsilon}Z_g) = KR'\; \Gamma_2(p,m^2,g\mu^{2\varepsilon})$.
\ice{
The RC $Z_2$ and $Z_{m^2}$ are  related with UV divergences of the bare
field propagator $D^B(p,m_0^2,g_0)$ in the following way:
\beq
D(p,m^2,g,\mu^2) =  \frac{1}{Z_2} \, D^B(p,m_0^2,g_0)
= \frac{1}{Z_2} \, \frac{1}{m_0^2 +p^2-
  \Gamma^0_2(p,m_0^2,g_0,\mu^2)} 
= \frac{1}{m^2 +p^2- \Gamma_2(p,m^2,g,\mu^2)}
{},
\eeq
where $D$ and $\Gamma_2$ stand for the (renormalized) propagator  and one
particle irreducible self-energy respectively.
They can be conveniently extracted from $\Gamma_2$ with the use of the
Bogoliubov-Parasiuk R-operation \cite{bog1957,bogshirk}:}
So RCs $Z_1$ and $Z_2$ can be conveniently extracted from $\Gamma_2$:
 \begin{equation}
Z_2 = 1+ \partial_{p^2}KR'\,{\Gamma}_2 (p,m^2,g,\mu), \ \ 
Z_1 = 1+ \partial_{m^2}KR'\,{\Gamma}_2 (p,m^2,g,\mu),
\end{equation}
where $R'$  is  the incomplete $R$-operation (which subtracts all proper UV subdivergences from a given  Feynman  amplitude
but does not  touch  its  UV divergence as  a whole) and $K$ stands for the operator extracting  the singular part of an $\vep$
expansion: 
\[
K \sum_i C_i \, \vep^i = \sum_{i < 0} C_i \,   \vep^i
{}.
\]
Renormalization constants $Z_i$ $i=1,2,3$ are known up to 5th-loop order
\cite{phi412,phi434,uniquenessK,phi456}. The aim of this paper is to extend the results of \cite{phi412}
by one more  order, that is  to  evaluate  analytically the sixth loop contribution to the anomalous dimension
$\gamma_{\vphi}$ and the  corresponding critical exponent $\eta$.

\ice{
 \begin{equation}
=1+\sum\limits_n (-u)^n\,\partial_{p^2} KR'\,{\Gamma}_2^{(n)} = 1 + \sum\limits_n\, (-u)^n\,
	\sum\limits_{\gamma\in\Gamma_2^{(n)}} s_\gamma \cdot \texttt{r}_\gamma\cdot \partial_{p^2} KR'(\gamma),
    \label{Z2}
\end{equation}
}

\section{RG calculations in MS-scheme: general  framework}
\label{sec:calcgen}

At present there are basically two different ways to perform the analytical RG
calculations at the multi-loop level. Both approaches make use of the
method of Infrared Rearrangement (IRR)~\cite{Vladimirov:1980zm,gpxt} in
order to make integral more suitable for analytical calculations by setting zero (possibly after a proper Taylor expansion) initial masses
and external momenta and introducing artificial ones. Both eventually employ the traditional integration
by parts method to compute the resulting Feynman integrals.
 
The first one \cite{Tarasov:1977ef,Kazakov:1979ik,Tarasov:1980au}
amounts to adding an artificial mass or an external momentum to a
properly chosen propagator of a given Feynman diagram before the
(formal) Taylor expansion in all masses (except for the artificial
one) and external momenta is made. The artificial external momentum
has to be introduced in such a way that all spurious infrared
divergences are softened away and the obtained Feynman integral is
calculable. In practice the condition of absence of the infrared
divergences leads to unnecessary complications and, in some cases,
even prevents from reduction to the simplest integrals. The problem
was solved by elaborating a special technique of subtraction of IR
divergences --- the $\wdt{R}$-operation \cite{ChS:R*,gssq,gvvq} which
we will discuss later.

\ice{
This technique succeeds in expressing the UV
counterterm of every L)-loop Feynman integral in terms of divergent and
finite parts of some (L-1)-loop massless propagators (which we will refer to as ``p-integrals'' in what  follows). 
}

In the second approach the infrared rearrangement is archived by inserting one
and the same auxiliary mass to {\em all} propagators 
\ice{(again after a formal expansion
in in all masses and external momenta)}
\cite{Misiak:1995zw,Chetyrkin:1997vx,vanRitbergen:1997va}.  After this no IR
divergences can ever appear. Next,  a proper expansion in all external momenta  and
particle masses (except the auxiliary one)  is to be 
performed. The resulting integrals are completely massive purely  vacuum integrals (tadpoles),
i.e. Feynman integrals without external momenta.
Note that the expansion in external momenta and masses (except for the
auxiliary one!) in both approaches is an unavoidable step if the (UV) RC we
are looking for is related to a non-logarithmically divergent Feynman
amplitude. It effectively reduces the quadratically (or even higher) UV
divergent amplitude  to the logarithmic one which opens the way to apply IRR to the 
latter. This is always possible within  dimensional regularization  and minimal subtractions schemes
(see, e. g.  \cite{Caswell:1981ek}).

Starting from $L=3$, $L$-massive tadpoles are getting significantly more
complicated for analytical evaluation than the L-loop vacuum integrals with
all but one massless  propagators.  As a result, the most advanced RG
calculations are being performed nowadays at the five loop level within the
first, ``massless'' approach (see, e.g. \cite{Baikov:2012zm,Eden:2012fe}).

Let us discuss now the current limits of the massless way of doing RG
calculations \ice{with $R^*$-operation and p-integrals} for  the example of a
logarithmically  divergent L-loop Feynman integral $\Fgamma$. We assume that all its UV 
subdivergences  are already known (the corresponding  Feynman (sub)-integrals will all have 
loop number strictly \mbox{ less then L)}.  Thus,  our aim  is to compute the  UV counterterm (we assume that
the original FI $\Fgamma$ is free from IR singularities)
\beq
Z_\gamma = -KR'\, \Fgamma
\nnb
 {}.
\eeq
The first two  steps are trivial: 
\\
(i) all (external momenta) and masses are set to zero;
\\ and\\ 
(ii)
the  integrand   of FI    $\Fgamma$  is modified by    introducing   a ``softening  factor''
\beq
\frac{p^2}{(p-q)^2}
{},
\label{mfactor}
\eeq
where the  momentum $p$ is the one flowing through an (arbitrary) internal line
$\ell$   (in principle,  one could equivalently use a combination
${p^2}/{(m_{aux}^2 + p^2)}$,  with $m_{aux}$ being an auxiliary  (non-zero) mass). 

The modified FI  $\Fgammaq$ is naturally represented as a convolution:
\beq
\Fgammaq = \int \frac{{\mathrm d} p  }{(2\,\pi)^D}\, 
\Fgammap (p) \, \frac{p^2}{(p-q)^2}
{},
\eeq
where the  (L-1)-loop p-integral\footnote{That is a massless integral,   depending
on only {\em one} external momenta  FI.}
\[
\Fgammap(p) =\, C_{\gamma'} (\vep)\, \frac{1}{(p^2)^{2+(L-1)\vep}}
\]
is obtained
by cutting the ``softened'' line $\ell$  in the the original diagram, that is 
$\gamma' = \gamma  \setminus  \ell$ .
Now, if by  a proper choice of $\ell$ the FI   $\Fgammaq$ is made free from  any IR divergences (such a choice is 
not always possible, see an example below) then 
\beq
Z_\gamma = -KR'\,\Fgammaq = -K\,  \Fgammaq + \dots
\label{dots}
{}.
\eeq 
Here dots stand for subtractions of UV {\em sub}divergences;  the corresponding
FI's all have loop number \mbox{strictly less then L} and, consequently, are  known according to our initial assumption. 
Thus,   the evaluation of $Z_\gamma$  amounts to the calculation  of the following  expression:
\bea
C_{\gamma'}(\vep)\, 
\int \frac{{\mathrm d} p  }{(2\,\pi)^D}\, 
\frac{1}{(p^2)^{2+(L-1)\vep}} \,\cdot\, \frac{p^2}{(p-q)^2}
&=& C_{\gamma'}(\vep)\,
( q^2)^{-L\ep} \ \ 
G(1+ (L-1)\,\vep,1)
\nnb
\\
&=&
C_{\gamma'}(\vep)\,
(q^2)^{-L\,\ep} \ \ \frac{1}{L\,\ep} \cdot (1+{\cal O}(\ep))
{}.
\label{basic:int}
\eea
or, equivalently, the function  $C_{\gamma'}(\vep)$ with accuracy ${\cal O}(\vep^0)$.
In the r.h.s of (\ref{basic:int}) we employ a convenient shortcut notation for
 a basic one loop p-integral \cite{phi412}:
\ice{\small K: I am not sure  about explicit definition of G-function below; depends on
whether we are  going to describe simple calculational examples later$\dots$  }
\beq
\int \frac{{d} \,p } {(2\,\pi)^d}\frac{1}{ (p^{2\alpha} )(q-p)^{2\beta}}\, =\,
\frac{(q^2)^{2 -\ep   -\al - \be }}{16\,\pi^2} 
\left( 
G(\alpha,\beta) = (4\pi)^\ep \, \frac{\Ga(\al +\be-2 +\ep)}{\Ga(\al)\Ga(\be)}
\frac{\Ga(2-\al-\ep)\,\Ga(2-\be-\ep)}{\Ga(4-\al - \be -2\ep)}
\right)
\label{G(al,be)}
{}.
\eeq

\begin{figure}[h]
\centering
\includegraphics{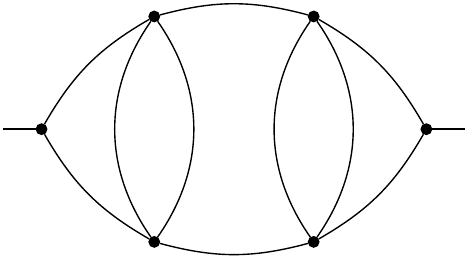}

\caption{No IR safe IRR (with one softened line) is  possible  for this graph.}
\label{fig:noIRsafeIRR}
\end{figure}

Unfortunately, the condition  of IR finitness  of the modified FI $\Fgammaq$ is rather restrictive, in many cases 
it prevents from a convenient  choice of the cut-line $\ell$ leading to a simpler for  calculation  (L-1)-loop
p-integral or even from the very possibility  of application of IRR to a diagram (see Fig.~\ref{fig:noIRsafeIRR}). 
The restriction can be lifted  completely with the use of $R^*$-operation which includes IR subtractions in addition to 
usual UV ones\ice{provided by  the R-operation}:
\beq
R^*  = R\cdot \widetilde{ R}
{}.
\eeq
Here $\widetilde{ R}$ stands for the  IR R-operation  which recursively subtracts  all IR singularities 
from a given (euclidean!) FI.  Thus, for  the  case of an arbitrary chosen line $\ell$   eq.  (\ref{dots})
assumes the form 
\beq
Z_\gamma = -KR'\,\wdt{R}\,  \Fgammaq =  -KR'\,\wdt{R}'\,  \Fgammaq   =   -K\,  \Fgammaq + \dots
\label{tildeZ}
{},
\eeq
Eqs. (\ref{tildeZ}) requires  a few comments. 

First, the $\wdt{R}'$ operation  is defined as   $\wdt{R}$   {\em without} the last IR subtraction corresponding
to  IR divergence of the FI  $\Fgammaq$ as a whole. The transition to $\wdt{R}'$ in  the middle  of (\ref{tildeZ}) 
is  perfectly legal as the presence of the modified propagator in  the FI $\Fgammaq$  ensures  the  superficial 
IR convergence of the  latter. 

Second,  the application of both $R'$ and $\wdt{R}'$ in (\ref{tildeZ}) is  a purely  algebraic procedure
as all UV and IR counterterms to be computed   can be algebraically  expressed\footnote{
We will not discuss in any detail the internal mechanics of  $\wdt{R}'$-operation (see in this connection
\cite{ChS:R*,Rstar_chet_mpi,acat14}.)} 
in terms of  (proper) UV counterterms of $\Fgamma$ (which are known according to our initial assumption).
As a  result, we again arrive at  a conclusion, that even for a generic choice of the cut-line $\ell$ 
the evaluation of  $Z_\gamma$ requires  knowledge  of the pole and finite parts of the (L-1)-loop p-integral
$\Fgammap$ (as well as some p-integrals with less number of loops).

\begin{figure}[h]
\centering
\includegraphics{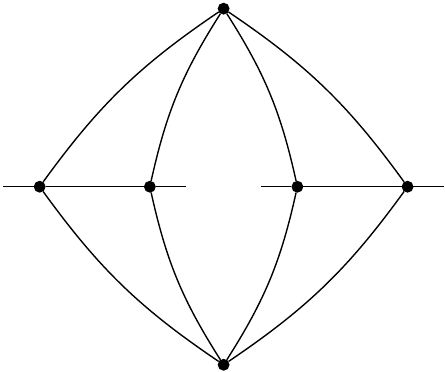}
\qquad\qquad
\includegraphics{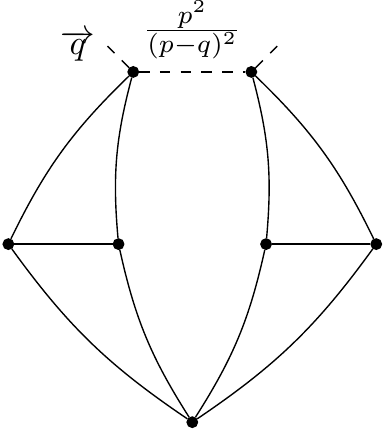}

\caption{IRR performed using vertex cut, dashed line represent corresponding softening factor.}
\label{fig:vertexcut}

\end{figure}

Third, given   a vertex  with more then three incident fields, it can  be  easily
transformed (cut)  into two vertices joined by  a new line with the corresponding propagator equal
identically 1 (see Fig.~\ref{fig:vertexcut}). This new line can also be used as a cut-one. We
will see in the next section that in many cases cutting  a vertex   
leads to  especially simple (in fact, factorizable) p-integrals.

Recently the state of the art of analytical calculation of p-integrals has established itself at the four loop level (for more details see \cite{Baikov:2015tea}),which means that 5-loop RG calculations are now feasible, while 6-loop calculations are still not possible in the general case. We will see in the next section why for a particular simple model of the scalar $\vphi^4$ theory the 5-loop barrier was taken more than thirty years ago and why these days even 6-loop level has got accessible.

\ice{ This technique succeeds in expressing the UV counterterm of every L-loop
  Feynman integral in terms of divergent and finite parts of some (L-1)-loop
  massless propagators (which we will refer to as ``p-integrals'' in what
  follows).  Thus, within the massless approach one should in principle the
  current limit of }

To summarize this section: given an L-loop completely massless 
\ice{is it correct? massless vacuum diagram has superficial IR divergence? Should we say that it not vacuum diagram,
but one softened by some auxiliary momenta?}vacuum diagram
$\Ga$ with zero (in four-dimensions) superficial index of the (UV) 
divergence of the corresponding formal FI $\FGamma$ the  use of
$R^*$ operation reduces the calculation of the UV counterterm $Z_{\Ga}$ to
evaluation of only one (L-1)-loop p-integral $\FGammap(p)$ obtained by
cutting an {arbitrary line}  $\ell$ from $\Ga$ (not counting p-integrals with
loop number less then L which should be computed for removing  UV and IR {\em
 sub}divergences from $\FGammaq$). 
The  final result for the UV  counterterm 
\[
KR'\,\wdt{R}' \Fgammaq
\] 
does not depend on the choice of the line $\ell$ which provides us with a
strong check of the correctness of the calculations.

\ice{Even  more important is the fact that}

\section{Calculation of TV-Reducible  diagrams}

The  main simplifying feature of the $\vphi^4$ model  comes from the  fact that
its only interaction vertex is composed of {\em four} scalar  fields.
As a result the variety of different ``topologies'' of FIs  
to be computed is strongly reduced with respect to, say, the $\vphi^3$ model.
This is well illustrated by the fact that the first analytical  four-loop RG  calculation
in the latter model have been performed very recently \cite{Gracey:2015tta}
(the four-loop RG-functions for the $\vphi^4$ model are known since 1979
\cite{Kazakov:1979ik}).

Different  cut-lines  lead  generically    not only  to different
(L-1)-loop  p-integrals:  a wisely chosen cut line could in many cases result
in    especially simple \ice{for calculation} p-integral. This happens if the original
vacuum graph $\Ga$  is TVR ({\em Two-Vertex-Reducible}). 
By definition, a 1PI vacuum graph  $\Ga$ belongs to a class of TVR  ones if  it is
possible to cut one of its lines or vertexes in such a way that the resulting
graph $\Ga \setminus \ell$  becomes  One-Vertex-Reducible (OVR), that is the
corresponding  FI $F_ {\Ga \setminus \ell}(p)$ can  be  presented  as a  product
of two p-integrals each with non-zero number of loops.

Thus, for a TVR graph the calculation of FI the $F_ {\Ga \setminus \ell}(p)$
amounts to computing  two p-integrals $F_{\ga_1}$ and $F_{\ga_2}$ with loop
numbers $L_1>0$ and $L_2>0$, $L_1+L_2= L-1$ respectively. This also means that
any  UV counterterm for every 6-loop FI $\FGamma$ (not necessarily
logarithmically divergent one) with $\Gamma$ being TVR is analytically calculable
provided  one knows the $\ep$ expansions of four-loop master p-integrals with
$\ep$ accuracy by {\rm one order more} then the one necessary for 5-loop
calculations\footnote{Actually, our current  calculation has {\em not} (accidentally?) 
required  this extra power of $\varepsilon$ for  four-loop master p-integrals. As a result
our final result for $\gamma_{\vphi}$ (see eq.~\ref{gphi.6loop}) does not include any irrational constants beyond those
appearing in general 5-loop RG calculations (for a detailed  discussion, see  \cite{baikov2010}).
}
 (and available from \cite{baikov2010}). Fortunately, this missing
power of $\ep$ 
\ice{actually we don't need them, for current calculations it
  is enough to know values from \cite{baikov2010}. may be we may say
  about it?  values from \cite{Lee:2011jt} required for beta function
  only.}
(and many more) have been all found in \cite{Lee:2011jt} for
the whole collection of 4-loop p-masters and confirmed in
\cite{Panzer:2013cha}.

\begin{figure}
\centering
\includegraphics{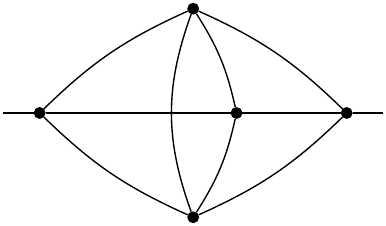}
\caption{
The only TVI diagram contributing  to the  field
  self-energy  at five loop.  
}
\label{fig:nonTVR5l}
\end{figure}

In fact, TVR graphs abound in the $\vphi^4$ model which is {the underlying
  reason} of the very possibility of the early 4 and 5-loop RG calculations as
well our current ability to perform the same calculations at the six  loop
level. Indeed, at three and  four  loops all diagrams contributing to the AD $\ga_2$ happen to be
TV-Reducible.  At  five loops all except  for {\em one} (see  Fig.~\ref{fig:nonTVR5l})
diagrams are also TV-Reducible. 

At  six  loop level    the situation is as follows: among  50  diagrams 
all are TVReducible except  for two.
To compute 48 TVR-diagrams  we have used  a (python) toolbox for  calculation
of UV countertems \cite{acat14} which allows to automate all operations on
Feynman diagrams, like infrared
rearrangement, $R^*$ operation as well as  IBP reduction (we have employed the
reduction rules generated by LiteRed \cite{litered}). The diagram-wise results are listed  in
Table \ref{tableG2} of   \ref{app1}. The table includes also the results for  {TV-Irreducible} 
diagrams whose treatment will be discussed in the next section.

\section{Calculation of  TV-Irreducible diagrams}
\begin{figure}[h]
\centering
\begin{subfigure}[b]{0.35\textwidth}
        \centering
\includegraphics{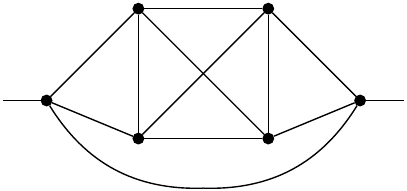}
\caption{ }
\label{nonTVRa}
    \end{subfigure}
\begin{subfigure}[b]{0.35\textwidth}
        \centering

\includegraphics{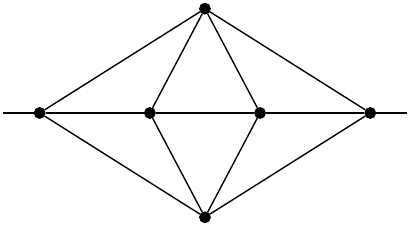}
\caption{ }
\label{nonTVRb}
    \end{subfigure}
\caption{
(a) and (b): TVI diagrams contributing   to the  field
  self-energy  at six loop level.  
\ice{
(a): the (only) TRI diagram contributing to  to the  field
  self-energy at five loops; (b) and (c): two TVI diagrams at six loops.
}
}
\label{fig:nonTVR}
\end{figure}

In six loops there are only two  TV-Irreducible diagrams pictured on  Fig.~\ref{fig:nonTVR}.
According to the general strategy  of IRR these diagrams  do  require the knowledge
of complicated (that is non-factorizable) 5-loop p-integrals for  their
evaluation. Below we describe  how  both diagrams  have been computed. 

\subsection{diagram (a)}
\label{subsec:diag1}

Diagram (a) (see Fig.\ref{nonTVRa}) has  quite a   special topology: it contains a line   connecting
both external  vertexes. In addition, it is quadratically divergent.  These
facts combined allow for rather simple calculation of the corresponding UV
\mbox{counterterm}.
First step is trivial as one among six loop integrations for diagram (a) can be
easily  done analytically (due to  a line connecting both external  vertexes) 
with the following result:
\begin{equation}
\begin{matrix}
\includegraphics{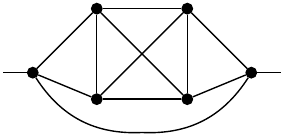}
\end{matrix}
= \, \frac{1}{16 \pi^2}\, G(1,5\,\ep)\,  
\begin{matrix}
\includegraphics{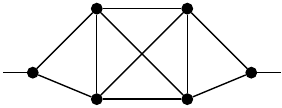}
\end{matrix}
\, ,{}\ \mbox{where (see eq. (\ref{G(al,be)}))}, \ G(1,5\,\ep)\,= \,  -\frac{5}{12} +{\cal O}(\ep).
\label{diag1dp}
\end{equation}
\ice{
with 
\beq
G(1,5\,\ep)\,= \,  -\frac{5}{12} +{\cal O}(\ep) 
\eeq
}
The fact that the first factor $G(1,5\,\ep)$ in r.h.s. of \eqref{diag1dp} is
of order ${\cal O}(\varepsilon^0)$ means that we need to know only pole part
of the second factor.  Pole part of this 5-loop p-integral is easy to compute
(see \ref{app2}).

\subsection{diagram (b)}
For the second diagram in Fig.~\ref{fig:nonTVR} we  need to calculate
the derivative with respect to $p$. This produces two terms (the line with an arrow stands for  $p_\mu/p^2$):
\begin{equation}
\frac{1}{2}(\partial_p)^2 KR'\left(\begin{matrix}
\includegraphics{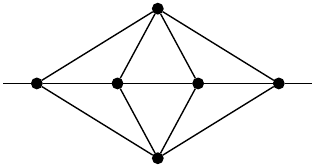}
\end{matrix}\right) = 2 KR'\left(\frac{4-d}{d}\begin{matrix}
\includegraphics{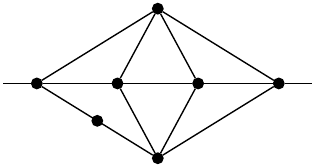}
\end{matrix}\right)+
2 KR'\left(\frac{4}{d}\begin{matrix}
\includegraphics{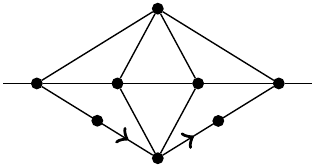}
\end{matrix}\right)\,,
\label{diag2dp}
\end{equation}
 The first diagram in r.h.s of
\eqref{diag2dp} can be calculated in the same way as first non TVR diagram in
Sec.~\ref{subsec:diag1}. The second one requires additional consideration.

First of all this diagram is logarithmically divergent  and primitive (i.e. contains no
subdivergences), so we can perform the following IR rearrangement:
\begin{equation}
KR'\left(\frac{4}{d}\begin{matrix}
\includegraphics{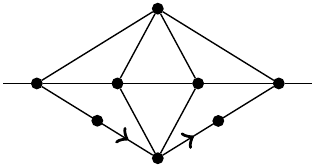}
\end{matrix}\right) = KR'\left(\frac{4}{d}\begin{matrix}
\includegraphics{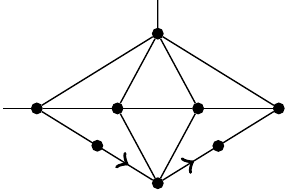}
\end{matrix} \right)
\end{equation}
For the latter diagram we can integrate out one loop using \eqref{basic:int}:
\begin{equation}
K\left(\frac{4}{d}\begin{matrix}
\includegraphics{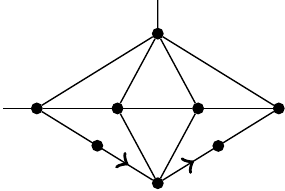}
\end{matrix} \right) = K\left(\frac{4}{d} G(1,1+5\varepsilon)\,
\begin{matrix}
\includegraphics{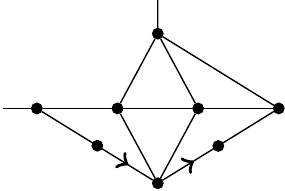}
\end{matrix}
\right)
\label{diag2IRR}
\end{equation}
We need the value of the  diagram in r.h.s of \eqref{diag2IRR} up to a constant
term only as the  corresponding factor there  is of order ${\cal O}(\varepsilon^{-1})$.  Because of
the fact that the diagram is finite (no  divergences at all) we need to
calculate only  the   leading (constant) term in its expansion in
$\varepsilon$. This can be done using
transition to the corresponding dual graph:
\vspace{1cm}
\begin{equation}
\left(\begin{matrix}
\includegraphics{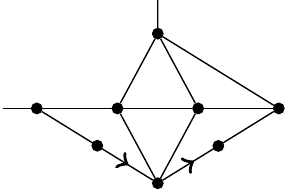}
\end{matrix}\right)_{\mbox{p-space}} = C\left(\begin{matrix}
\includegraphics{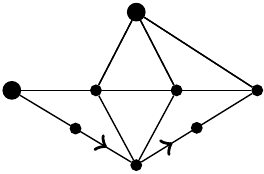}
\end{matrix}\right)_{\mbox{x-space}}=C\left(
\begin{matrix}
\includegraphics{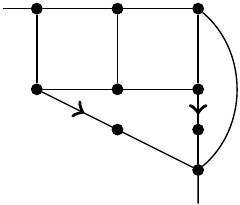}
\end{matrix}
\right)_{\mbox{p-space}}
\label{diag2dual}
\end{equation}
\vspace{1cm}

It should be noted the the x-space propagators (middle and the right  diagrams  in \eqref{diag2dual})
have a non-standard  $\varepsilon$ dependence, viz. $1/(x_1 -x_2)^{2(1-\varepsilon)}$.
Fortunately,  as far as we are looking only for leading (constant)
contribution we can consider standard propagators $1/(x_1 -x_2)^2$.
Now, the diagram in r.h.s of \eqref{diag2dual} has only 4 loops and can be
calculated using the standard 4-loop IBP reduction. The fact that transition to
the dual graph can lower the number of loops is another simplifying feature of
the $\varphi^4$ model. Interestingly, the 5-loop TVI diagram on Fig.~\ref{fig:nonTVR5l}  can  be
also easily performed in the same way\footnote{Originally the diagram was
analytically computed in \cite{chet3} with  a series of ad-hoc non-obvious tricks.}.

\section{Results and discussion \label{res_dis}}

After adding diagram-wise results of Table~\ref {tableG2} and known five  loop
results \cite{phi456} we arrive  at the following expression for 
the  anomalous dimension of field $\gamma_\varphi$ to the six loop level:
\begin{equation}
\begin{split}
\gamma_\varphi(g)= & \frac{\ccu{2}(n+2)}{36} - \bigg[ 8 + n \bigg]\frac{\ccu{3}(n+2)}{432} +  \bigg[ 500+90 \; n-5 \; n^{2} \bigg]\frac{\ccu{4}(n+2)}{5184} + \bigg[ -77056 + 8832 \; \zeta_{3} - 25344 \; \zeta_{4}+
\\ + &
\left(-22752 + 3072 \; \zeta_{3} - 5760 \; \zeta_{4}\right) \; n+\left(-296 - 288 \; \zeta_{3}\right) \; n^{2}+\left(-39 + 48 \; \zeta_{3}\right) \; n^{3} \bigg]\frac{\ccu{5}(n+2)}{186624} + \\ +  
 & \bigg[ 1410544 + 1190400 \; \zeta_{6} + 297472 \; \zeta_{3} - 833536 \; \zeta_{5} - 95232 \; \zeta_{3}^{2} + 619776 \; \zeta_{4}+
 \\ + &
 \left(549104 + 352000 \; \zeta_{6} + 69888 \; \zeta_{3} - 293632 \; \zeta_{5} - 28160 \; \zeta_{3}^{2} + 215808 \; \zeta_{4}\right) \; n+
 \\ + &
 \left(30184 + 12800 \; \zeta_{6} + 14976 \; \zeta_{3} - 23680 \; \zeta_{5} - 1024 \; \zeta_{3}^{2} + 15744 \; \zeta_{4}\right) \; n^{2}+\left(-794 + 96 \; \zeta_{4}\right) \; n^{3}+
 \\ + &
 \left(-29 - 16 \; \zeta_{3} + 48 \; \zeta_{4}\right) \; n^{4} \bigg]\frac{\ccu{6}(n+2)}{746496}
{}.
\end{split}
\label{gphi.6loop}
\end{equation}
Substituting $g_*$ calculated in 5 loop approximation (see
e.g. \cite{Vasiliev}) into the anomalous dimension $\gamma_2 = 2\,
\gamma_\varphi$ we  obtain the critical exponent $\eta$ up to ${\cal   O}(\varepsilon^7)$:
\begin{equation}
\begin{split}
\eta(\varepsilon)= & \frac{(2\epsilon)^{2}}{2}\frac{(n+2)}{(n+8)^2} +\bigg[ 272+56 \; n- n^{2} \bigg]\frac{(2\epsilon)^{3}}{8}\frac{(n+2)}{(n+8)^4} +  
  \bigg[ 46144 - 67584 \; \zeta_{3}+\left(17920 -  23808 \; \zeta_{3}\right) \; n + 
  \\ + &
  \left(1124 - 1920 \; \zeta_{3}\right) \; n^{2}-230 \; n^{3}-5 \; n^{4} \bigg]\frac{(2\epsilon)^{4}}{32}\frac{(n+2)}{(n+8)^6} +  \bigg[ 5655552 + 60948480 \; \zeta_{5} - 21921792 \; \zeta_{3} - 
  \\ - & 
  12976128 \; \zeta_{4}+
  \left(2912768 + 33259520 \; \zeta_{5} - 11530240 \; \zeta_{3} - 7815168 \; \zeta_{4}\right) \; n+
    \\ + &
    \left(262528 + 6113280 \; \zeta_{5} - 1244160 \; \zeta_{3} - 1714176 \; \zeta_{4}\right) \; n^{2}+\left(-121472 + 445440 \; \zeta_{5} + 137984 \; \zeta_{3} - 163584 \; \zeta_{4}\right) \; n^{3}+
      \\ + &
      \left(-27620 + 10240 \; \zeta_{5} + 20800 \; \zeta_{3} - 5760 \;
        \zeta_{4}\right) \; n^{4}+\left(-946 + 288 \; \zeta_{3}\right) \;
      n^{5}+\left(-13 + 16 \; \zeta_{3}\right) \; n^{6}
      \bigg]\frac{(2\epsilon)^{5}}{128}\frac{(n+2)}{(n+8)^8} +
\end{split}
\label{etae6}
\eeq
\beq
\begin{split}
\phantom{\eta(\varepsilon)=} &+ \bigg[ 565354496 - 60808495104 \; \zeta_{7} + 19134414848 \; \zeta_{5} + 19503513600 \; \zeta_{6} - 5485101056 \; \zeta_{3} + 5036310528 \; \zeta_{3}^{2} -
   \\ - &
   4208984064 \; \zeta_{4}+\left(323108864 - 44652625920 \; \zeta_{7} + 13118341120 \; \zeta_{5} + 15518924800 \; \zeta_{6} - 3681222656 \; \zeta_{3} +
     \right.\\ + & \left.
     4007919616 \; \zeta_{3}^{2} - 3266052096 \; \zeta_{4}\right) \; n+\left(8413184 - 12662415360 \; \zeta_{7} + 2504949760 \; \zeta_{5} + 4921753600 \; \zeta_{6} 
     \right.\\ - & \left.     
     - 533012480 \; \zeta_{3} + 1142210560 \; \zeta_{3}^{2} - 858095616 \; \zeta_{4}\right) \; n^{2}+\left(-45721600 - 1749888000 \; \zeta_{7} - 84449280 \; \zeta_{5} +
    \right.\\ + & \left.
    797900800 \; \zeta_{6} + 131311616 \; \zeta_{3} + 144695296 \; \zeta_{3}^{2} - 67817472 \; \zeta_{4}\right) \; n^{3}+\left(-17128928 - 118540800 \; \zeta_{7} - 71895040 \; \zeta_{5} + 
     \right.\\ + & \left.     
     69478400 \; \zeta_{6} + 40585984 \; \zeta_{3} + 8321024 \; \zeta_{3}^{2} + 6884352 \; \zeta_{4}\right) \; n^{4}+\left(-2460768 - 3161088 \; \zeta_{7} - 6955264 \; \zeta_{5} +
     \right.\\ + & \left.     
     3046400 \; \zeta_{6} + 2822400 \; \zeta_{3} + 250880 \; \zeta_{3}^{2} + 1467648 \; \zeta_{4}\right) \; n^{5}+\left(-110512 - 195200 \; \zeta_{5} + 51200 \; \zeta_{6} + 36096 \; \zeta_{3} +
     \right.\\ + & \left.
     8192 \; \zeta_{3}^{2} + 79296 \; \zeta_{4}\right) \; n^{6}+\left(-2748 + 2656 \; \zeta_{3} + 1632 \; \zeta_{4}\right) \; n^{7}+\left(-29 - 16 \; \zeta_{3} + 48 \; \zeta_{4}\right) \; n^{8} \bigg]\frac{(2\epsilon)^{6}}{512}\frac{(n+2)}{(n+8)^{10}}
{}.
\end{split}
\nonumber
\end{equation}

For $n=1$ the  anomalous dimension $\gamma_\varphi$ and the exponent $\eta$ assume the form
\bea
\gamma_\varphi &=&  \frac{1}{12}g^{2}  -\frac{1}{16}g^{3} +  
   \frac{65}{192}g^{4} +  \bigg[ -3709 - 1152 \; \zeta_{4} + 432 \; \zeta_{3}
   \bigg]\frac{g^{5}}{2304} +
\label{gamma.phi.n-=1}
 \\ &+  
 & \mathbf{\bigg[ 73667 + 31536 \; \zeta_{4} - 4608 \; \zeta_{3}^{2} + 57600 \; \zeta_{6} - 42624 \; \zeta_{5} + 14160 \; \zeta_{3} \bigg]\frac{g^{6}}{9216}} + {\cal O}(u^{7})\nonumber \\
&=&0.0833 \ccu2 -0.0625 \ccu3 + 0.3385 \ccu4 -1.9255
\ccu5 +\mathbf{14.383} \ccu6  +{\cal O}(\ccu7),
\nonumber
\\ 
\eta&=&\frac{2}{27}\varepsilon^2+\frac{109}{729}\varepsilon^3+\left(\frac{7217}{39366}-\frac{64}{243}
  \zeta_3\right)\varepsilon^4+\left(\frac{321511}{2125764}-\frac{32}{81}
  \zeta_4-\frac{1316}{2187} \zeta_3+\frac{1280}{729} \zeta_5\right
)\varepsilon^5+
\\
&+&\mathbf{\bigg(\frac{3421613}{38263752}-\frac{3136}{243}
 \zeta_7+\frac{73232}{19683} \zeta_5
-\frac{181462}{177147} \zeta_3+\frac{3200}{729} \zeta_6+\frac{2432}{2187} 
 \zeta_3^{2}-\frac{658}{729} \zeta_4\bigg
 )\varepsilon^6}+\mathcal{O}(\varepsilon^7) = 
\nonumber
\\ 
 &=& 0.074074\,
 \varepsilon^2+0.149520\,\varepsilon^3-0.133260\,\varepsilon^4+0.821006\,
 \varepsilon^5-\mathbf{5.201449}\,\varepsilon^6+\mathcal{O}(\varepsilon^7)
{}.
\eea

We perform various consistency checks of our results.  First of all,
the finitness of $\gamma_\vphi$ as found from (\ref{g_phi}) at 6 loop
ensures   the   correctness of high order poles in $\varepsilon$ in the RC $Z_\vphi$.
The first pole in $\varepsilon$
(which actually contributes to $\gamma_\varphi$ and $\eta$) cannot be checked
in such a way. Fortunately, there is  a self-consistency test which is
sensitive to the structure of the first pole. It is
based on the  known results of $1/n$-expansion for critical exponent $\eta$. 
The expansion is currently available  up to $1/n^3$ term
\cite{oneovern3,Vasiliev}.
The coefficients of this
expansion are exact functions of  $\varepsilon$, on the other hand coefficients
of $\varepsilon$-expansion of critical exponent $\eta$ are exact functions on
$n$. Expanding both functions in  $\varepsilon$ and $1/n$ respectively we will
obtain double expansion in $\varepsilon$ and $1/n$ which must coincide up to given
($\varepsilon^6,\; 1/n^3$) order. From these  expansions we can derive 3
independent relations on linear combinations of the coefficients at first pole
in $\varepsilon$ of the six loop diagrams. Moreover, a relation that originates
from the term of order $1/n^3$ includes all graphs from 
\ref{tableG2}.
All three relations are indeed in agreement with our results (more details 
can  be found in Appendix \ref{error_in_29}).

For some selected diagrams we have also  performed additional
numerical checks  using the sector decomposition technique (see, e.g \cite{heinrich}).

In papers \cite{Kazakov:1980rd,Kazakov:1978ey} the method of a resummation of the asymptotic series was proposed.  This method combines an assumption  about asymptotic of beta function at $g\to\infty$ and  the available information  about higher order asymptotic \cite{Lipatov} via  a Borel transformation with conformal mapping. 
It was shown that for the series where asymptotic $g\to\infty$ is known, most accurate values (after resummation of the finite part of the series) are obtained if parameter $\nu$ (additional parameter which defines the behavior of the resummed series at $g\to\infty$) is chosen in accordance with $g\to\infty$ asymptotic. More over in this case the contribution of high order terms gets  minimized.

For the $\varphi^4$ model the asymptotic behavior at $g\to\infty$ is not known, so
authors of \cite{Kazakov:1980rd,Kazakov:1978ey} used the  criterion of minimization of
the contribution of the high order terms as a way to determine the  correct value
of the \mbox{parameter $\nu$}.  They found that for the case of the beta-function of
the $\varphi^4$ model   it should lie within the range
$1.7<\nu<2.2$, commonly  the  value $\nu=2$ is taken.
 
Furthermore, if we perform such a resummation procedure for a given number of
loops $L$ and then expand back the series obtained after conformal mapping
procedure up to the next, $(L+1)$-loop  order, then this term may be considered as a
prediction  for the $(L+1)$-loop contribution because of the minimization of high order contributions
we have discussed above. In particular, the prediction of
\cite{Kazakov:1980rd,Kazakov:1978ey} 
for the 5-loop term in the beta-function
happened to be $1404.3$ while the direct calculations
\cite{phi434,uniquenessK,phi456} (published a year later) produced the value
1424.28, which is different from the prediction only by a minute $1.5\%$.

We apply the  same procedure to $\gamma_\varphi$ with $n=1$ (see eq.~(\ref{gamma.phi.n-=1})).
\ice{
where $\gamma_\varphi$ looks as follows:
\begin{equation}
\gamma_\varphi(g) =  0.0833 \ccu2 -0.0625 \ccu3 + 0.3385 \ccu4 -1.9255\ccu5 +14.383\ccu6+{\cal O}(\ccu7).
\end{equation}
}
Using $\nu=3$ and performing the same steps for terms up to 5 loops we arrive
to the following predictions for the 6-loop term
\begin{equation}
\gamma_\varphi^{P5}(g) =  0.0833 \ccu2 -0.0625 \ccu3 + 0.3385 \ccu4 -1.9255 \ccu5 \mathbf{+14.316 \ccu6}+{\cal O}(\ccu7)
{},
\end{equation}
which is only by $0.5\%$ smaller than calculated in the present  work.
If we repeat  the same procedure starting from  6 loops we can make a
prediction for the 7 loop contribution to the  field anomalous dimension.
\begin{equation}
\gamma_\varphi^{P6}(g) =  0.0833 \ccu2 -0.0625 \ccu3 + 0.3385 \ccu4 -1.9255 \ccu5 +14.383 \ccu6\mathbf{-127.29\ccu7}+{\cal O}(\ccu8)
{}.
\nonumber
\end{equation}\
\ice{
Of course,  for a  better  estimation of  high order coefficients one needs to
determine asymptotic behavior o large coupling constant.
}

If one perform a resummation of the $\gamma_\varphi(g)$ at $g=g_*$ 
(where $g_*$ is a first positive zero of the resummed beta-function),
one can obtain estimations for the Fisher exponent $\eta$ for 
different numbers of loops taken into account in $\beta(g)$ and $\gamma_\varphi(g)$:
\begin{table}[h!]
\centering
\begin{tabular}{c| c c c| c c c| c c|c} 
\hline
Loops $\beta/\gamma_\varphi$& 3/4& 3/5 & 3/6& 4/4 & 4/5 & 4/6 & 5/5 & 5/6 &est.\\[0.5ex] 
 \hline
$D=2$ &0.1716 & 0.1818& 0.1827& 0.2211& 0.2365& 0.2379& 0.2263& 0.2276 & 0.25\\[0.5ex] 
$D=3$ &0.03201& 0.03256& 0.03260& 0.03557& 0.03624& 0.03629& 0.03577& 0.03581 & 0.03601\\
\hline
 \end{tabular}
\caption{Resummation result for the Fisher exponent $\eta$ for different number of loops taken into account}
\label{table:resummed}
\end{table}

The column in Table \ref{table:resummed} marked as 'est.'  is an
estimated value for this model($n=1$). For  two dimensional model it
corresponds to the Onsager exact solution, for three dimensional case
it corresponds to a combination of  of the results of the high temperature
expansion~(HT) and the Monte-Carlo simulations~(MC) made in
\cite{critreview}.  One can see that results of resummation for the  3D
model are very close to HT and MC results ($\sim 0.5\%$). For the  2D model results are
also in reasonable agreement with the  Onsager exact solution but still far from it ($\sim 5-10\%$).
This effect may be explained by large value of the
expansion parameter $\epsilon=1$. Also one can see from the table that
most valuable impact on the value of the Fisher exponent is given by
the value of the fixed point (i.e. beta function). This fact may serve
as an additional argument to compute 6-loop beta function\cite{beta6},
of course, for the 3D model one may expect swing around the HT/MC value, but
for the 2D model, due to significant impact of the 5-loop beta
function(comparing to the 4-loop one) we still can't expect reasonable
result for the Fisher exponent.

\section{Conclusions}

We have described a completely analytical calculation of the field
anomalous dimension $\gamma_{\vphi}$ and the critical exponent $\eta$
for the $O(n)$-symmetric $\vphi^4$ model at the six loop level. The
calculation has proved to be possible due to a combination of the
method of IRR based on the heavy use of the $R^*$-operation and recent
advances in computing master four-loop massless p-integrals as well as
due to a special feature of the $\vphi^4$ theory: the overwhelming
number of diagrams appearing at 4- and 5-loops happen to be  Two Vertex
Reducible ones.

We successfully compare our result for $\gamma_{\vphi}$ with $n=1$ with the
predictions based on the method of the Borel transform followed by a conformal
mapping. 
\ice{
It is found that the relative accuracy of the predictions is increased
with number of loops \todo{actually not. beta function does not show it. drop?}
}

Our diagram-wise results for all six loop contributions to $Z_2$ (together with some
auxiliary information
\ice{ like a compilation of all available  lower order
results for separate contributions to the renormalization
constants of the  $\varphi^4$ model)
} 
are  available (in
computer-readable form) in  \ \ 
\url{http://www.ttp.kit.edu/Progdata/ttp15/ttp15-046/}

They are also appended to the \TeX-file of the present paper.

\vspace{4mm}

\noindent
{\bf Acknowledgements}
\vspace{4mm}

We thank J.~H.~K\"uhn for an attentive reading of the
manuscript and L.~Ts.~Adzhemyan, D.~I.~Kazakov, R.~N.~Lee for fruitful
discussions.  

This work was supported by the Deutsche Forschungsgemeinschaft through CH 1479/1-1 (K.G.~Ch.).  We acknowledge Saint-Petersburg State University for a
research grant 11.38.185.2014 (D.V.~B. and M.V.~K.).  We also thank
Resource Center ``Computer Center of SPbU'' for providing computational
resources.

\def\appendixname{Appendix }

\appendix

\section{Diagramwise results for  6-loop contributions to $Z_2$   \label{app1}}

Tables \ref{tableG2} and \ref{tabler} display results for all fifty self-energy diagrams contributing
to RC $Z_2$.  For brevity we have used the so-called Nickel index (NI) which allows
for a short and  concise  description  of a given diagram
\cite{nickel,graphstate}.   

\begin{figure}
\centering
\includegraphics{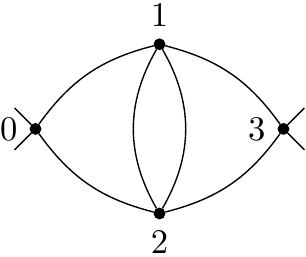}
\caption{Graph that corresponds to Nickel index (NI) equal to $ee12|223|3|ee|$.}
\label{fig:ni}
\end{figure}

Generally speaking Nickel index is a list of graph edges  written for some canonical
vertex ordering. The canonical vertex ordering ensures that two isomorphic
graphs have  equal Nickel indices.  
For example, consider Nickel index $'ee12|223|3|ee|'$: vertical lines split the
NI on sections, each section corresponds to the one of the vertices. Vertices
are assumed to be labeled from 0 (up to 3 for this graph), each section
describes graph edges connected to this vertex, i.e. vertex 0 has two external
($e$) edges and edges to vertices 1 and 2. Next section lists edges connected
to vertex 1 (except ones that connected to the vertex 0): two edges to vertex
2 and edge to 3. Third section lists edges connected to vertex 2 (except ones
connected to 0 and 1) and so on...  Drawing graph in such a way we arrive to
the diagram on Fig.~\ref{fig:ni}.

Construction of the NI from the graph is a bit more complicated task: one need
to take all possible graph labeling, for each labeling write a Nickel notation
described above, and then choose minimal(in some sense) notation as
NI. Luckily this procedure can be optimized to avoid $n!$ growth (see
\cite{graphstate}).

Every row in 
\ref{tableG2} describes a contribution of a diagram $\gamma $ with NI $
\mbox{NI}(\gamma)$ to $Z_2$ as  a product of three factors, namely,    $s_\gamma$ (a
symmetry factor),  $\texttt{r}_\gamma$ (an 
additional structure factor for n-component $O(n)$-symmetric
$\varphi^4$-model {in terms of polynomials} given in 
\ref{tabler}) and, finally, the very counterterm $\partial_{p^2}KR'\gamma$.

\newpage

\def\arraystretch{1.7}

\begin{longtable}{| l | c | c | c | l | }
\caption{Values of the six loop graphs contributing to $Z_2$}\label{tableG2}\\
\hline
N & NI$(\gamma)$ & $s_\gamma$ &$\texttt{r}_\gamma$ & $\qquad \partial_{p^2}  KR'\gamma$\\ \hline\hline
1 &
$e112|23|34|45|55|e|$ &
$ 1/4 $ &
$ r_{1} \; r_{10} $ &
$ -\frac{1}{90} \; \varepsilon^{-5}+\frac{13}{180} \; \varepsilon^{-4}-\frac{13}{45} \; \varepsilon^{-3}+\frac{133}{180} \; \varepsilon^{-2}-\frac{4}{3} \; \varepsilon^{-1} $ 
\\ \hline 
2 &
$e112|23|34|55|e55||$ &
$ 1/4 $ &
$ r_{1} \; r_{13} $ &
$ -\frac{1}{48} \; \varepsilon^{-5}+\frac{121}{1440} \; \varepsilon^{-4}-\frac{11}{64} \; \varepsilon^{-3}+\left(\frac{289}{5760} + \frac{13}{120} \; \zeta_{3}\right) \; \varepsilon^{-2}+\left(\frac{5809}{11520} - \frac{1}{40} \; \zeta_{4} - \frac{53}{240} \; \zeta_{3}\right) \; \varepsilon^{-1} $ 
\\ \hline 
3 &
$e112|23|34|e5|555||$ &
$ 1/6 $ &
$ r_{1}^{2} \; r_{3} $ &
$ -\frac{31}{2880} \; \varepsilon^{-4}+\frac{191}{2880} \; \varepsilon^{-3}-\frac{47}{256} \; \varepsilon^{-2}+\frac{1675}{4608} \; \varepsilon^{-1} $ 
\\ \hline 
4 &
$e112|23|44|455|5|e|$ &
$ 1/8 $ &
$ r_{1} \; r_{11} $ &
$ -\frac{1}{40} \; \varepsilon^{-5}+\frac{7}{80} \; \varepsilon^{-4}-\frac{67}{480} \; \varepsilon^{-3}+\left(\frac{29}{960} + \frac{1}{20} \; \zeta_{3}\right) \; \varepsilon^{-2}+\left(-\frac{49}{640} + \frac{3}{40} \; \zeta_{4} - \frac{43}{120} \; \zeta_{3}\right) \; \varepsilon^{-1} $ 
\\ \hline 
5 &
$e112|23|44|555|e5||$ &
$ 1/12 $ &
$ r_{1}^{2} \; r_{3} $ &
$ -\frac{1}{64} \; \varepsilon^{-4}+\frac{3}{64} \; \varepsilon^{-3}-\frac{19}{3840} \; \varepsilon^{-2}+\left(\frac{707}{7680} + \frac{1}{30} \; \zeta_{3}\right) \; \varepsilon^{-1} $ 
\\ \hline 
6 &
$e112|23|44|e55|55||$ &
$ 1/8 $ &
$ r_{1} \; r_{9} $ &
$ -\frac{11}{240} \; \varepsilon^{-5}+\frac{41}{480} \; \varepsilon^{-4}-\frac{23}{960} \; \varepsilon^{-3}+\left(-\frac{11}{384} - \frac{1}{24} \; \zeta_{3}\right) \; \varepsilon^{-2}+\left(-\frac{187}{768} + \frac{1}{8} \; \zeta_{4} + \frac{1}{48} \; \zeta_{3}\right) \; \varepsilon^{-1} $ 
\\ \hline 
7 &
$e112|23|45|445|5|e|$ &
$ 1/2 $ &
$ r_{1} \; r_{10} $ &
$ -\frac{7}{720} \; \varepsilon^{-5}+\frac{17}{288} \; \varepsilon^{-4}-\frac{563}{2880} \; \varepsilon^{-3}+\left(\frac{2269}{5760} - \frac{13}{60} \; \zeta_{3}\right) \; \varepsilon^{-2}+\left(-\frac{497}{3840} + \frac{1}{20} \; \zeta_{4} + \frac{9}{40} \; \zeta_{3}\right) \; \varepsilon^{-1} $ 
\\ \hline 
8 &
$e112|23|45|e45|55||$ &
$ 1/4 $ &
$ r_{1} \; r_{11} $ &
$ -\frac{11}{720} \; \varepsilon^{-5}+\frac{103}{1440} \; \varepsilon^{-4}-\frac{127}{960} \; \varepsilon^{-3}+\left(\frac{31}{1152} + \frac{1}{5} \; \zeta_{3}\right) \; \varepsilon^{-2}+\left(\frac{2843}{11520} - \frac{3}{40} \; \zeta_{4} - \frac{3}{10} \; \zeta_{3}\right) \; \varepsilon^{-1} $ 
\\ \hline 
9 &
$e112|23|e4|455|55||$ &
$ 1/4 $ &
$ r_{1}^{2} \; r_{2}^{2} $ &
$ -\frac{7}{720} \; \varepsilon^{-4}+\frac{37}{720} \; \varepsilon^{-3}-\frac{307}{2880} \; \varepsilon^{-2}+\left(-\frac{1}{240} + \frac{1}{120} \; \zeta_{3}\right) \; \varepsilon^{-1} $ 
\\ \hline 
10 &
$e112|33|344|5|55|e|$ &
$ 1/16 $ &
$ r_{1} \; r_{14} $ &
$ -\frac{7}{180} \; \varepsilon^{-5}+\frac{3}{40} \; \varepsilon^{-4}+\frac{7}{720} \; \varepsilon^{-3}+\left(-\frac{59}{480} + \frac{1}{10} \; \zeta_{3}\right) \; \varepsilon^{-2}+\left(\frac{959}{2880} + \frac{3}{20} \; \zeta_{4} - \frac{1}{20} \; \zeta_{3}\right) \; \varepsilon^{-1} $ 
\\ \hline 
11 &
$e112|33|444|55|5|e|$ &
$ 1/48 $ &
$ r_{1}^{2} \; r_{4} $ &
$ -\frac{1}{40} \; \varepsilon^{-4}+\frac{13}{320} \; \varepsilon^{-3}+\frac{29}{320} \; \varepsilon^{-2}+\left(\frac{221}{1280} - \frac{7}{80} \; \zeta_{3}\right) \; \varepsilon^{-1} $ 
\\ \hline 
12 &
$e112|33|445|45|5|e|$ &
$ 1/4 $ &
$ r_{1} \; r_{13} $ &
$ -\frac{13}{720} \; \varepsilon^{-5}+\frac{103}{1440} \; \varepsilon^{-4}-\frac{377}{2880} \; \varepsilon^{-3}+\left(\frac{155}{1152} - \frac{19}{120} \; \zeta_{3}\right) \; \varepsilon^{-2}+\left(\frac{703}{1280} - \frac{1}{20} \; \zeta_{4} - \frac{1}{48} \; \zeta_{3}\right) \; \varepsilon^{-1} $ 
\\ \hline 
13 &
$e112|33|445|e5|55||$ &
$ 1/16 $ &
$ r_{1}^{2} \; r_{2}^{2} $ &
$ -\frac{1}{72} \; \varepsilon^{-4}+\frac{1}{18} \; \varepsilon^{-3}-\frac{7}{288} \; \varepsilon^{-2}-\frac{11}{24} \; \varepsilon^{-1} $ 
\\ \hline 
14 &
$e112|33|e34|5|555||$ &
$ 1/12 $ &
$ r_{1}^{2} \; r_{4} $ &
$ -\frac{13}{720} \; \varepsilon^{-4}+\frac{197}{2880} \; \varepsilon^{-3}-\frac{1}{48} \; \varepsilon^{-2}+\left(-\frac{223}{2304} - \frac{1}{48} \; \zeta_{3}\right) \; \varepsilon^{-1} $ 
\\ \hline 
15 &
$e112|33|e44|55|55||$ &
$ 1/32 $ &
$ r_{1} \; r_{15} $ &
$ -\frac{1}{12} \; \varepsilon^{-5}+\frac{1}{24} \; \varepsilon^{-4}+\frac{5}{48} \; \varepsilon^{-3}+\left(\frac{13}{96} - \frac{1}{6} \; \zeta_{3}\right) \; \varepsilon^{-2}+\left(\frac{29}{192} - \frac{1}{4} \; \zeta_{4} + \frac{1}{12} \; \zeta_{3}\right) \; \varepsilon^{-1} $ 
\\ \hline 
16 &
$e112|33|e45|45|55||$ &
$ 1/8 $ &
$ r_{1} \; r_{9} $ &
$ -\frac{1}{36} \; \varepsilon^{-5}+\frac{31}{360} \; \varepsilon^{-4}-\frac{13}{240} \; \varepsilon^{-3}+\left(-\frac{11}{288} - \frac{1}{60} \; \zeta_{3}\right) \; \varepsilon^{-2}+\left(-\frac{511}{2880} - \frac{1}{40} \; \zeta_{4} + \frac{23}{120} \; \zeta_{3}\right) \; \varepsilon^{-1} $ 
\\ \hline 
17 &
$e112|34|334|5|55|e|$ &
$ 1/8 $ &
$ r_{1} \; r_{10} $ &
$ -\frac{1}{72} \; \varepsilon^{-5}+\frac{13}{240} \; \varepsilon^{-4}-\frac{17}{288} \; \varepsilon^{-3}+\left(\frac{5}{192} - \frac{1}{10} \; \zeta_{3}\right) \; \varepsilon^{-2}+\left(-\frac{341}{5760} - \frac{3}{20} \; \zeta_{4} + \frac{7}{60} \; \zeta_{3}\right) \; \varepsilon^{-1} $ 
\\ \hline 
18 &
$e112|34|335|4|55|e|$ &
$ 1/8 $ &
$ r_{1} \; r_{12} $ &
$ -\frac{1}{72} \; \varepsilon^{-5}+\frac{13}{240} \; \varepsilon^{-4}-\frac{17}{288} \; \varepsilon^{-3}+\left(\frac{5}{192} + \frac{1}{15} \; \zeta_{3}\right) \; \varepsilon^{-2}+\left(-\frac{341}{5760} - \frac{11}{40} \; \zeta_{4} + \frac{9}{20} \; \zeta_{3}\right) \; \varepsilon^{-1} $ 
\\ \hline 
19 &
$e112|34|335|5|e55||$ &
$ 1/4 $ &
$ r_{1} \; r_{11} $ &
$ -\frac{7}{360} \; \varepsilon^{-5}+\frac{1}{15} \; \varepsilon^{-4}-\frac{37}{720} \; \varepsilon^{-3}+\left(-\frac{17}{480} - \frac{1}{60} \; \zeta_{3}\right) \; \varepsilon^{-2}+\left(\frac{553}{2880} - \frac{1}{40} \; \zeta_{4} - \frac{19}{120} \; \zeta_{3}\right) \; \varepsilon^{-1} $ 
\\ \hline 
20 &
$e112|34|335|e|555||$ &
$ 1/24 $ &
$ r_{1}^{2} \; r_{3} $ &
$ -\frac{7}{480} \; \varepsilon^{-4}+\frac{11}{240} \; \varepsilon^{-3}+\frac{7}{384} \; \varepsilon^{-2}+\frac{3}{256} \; \varepsilon^{-1} $ 
\\ \hline 
21 &
$e112|34|345|45|5|e|$ &
$ 1/2 $ &
$ r_{1} \; r_{2} \; r_{3} $ &
$ -\frac{4}{15} \; \zeta_{3} \; \varepsilon^{-3}+\left(\frac{1}{10} \; \zeta_{4} + \frac{19}{30} \; \zeta_{3}\right) \; \varepsilon^{-2}+\left(-\frac{17}{40} \; \zeta_{4} - \frac{13}{10} \; \zeta_{3} + \frac{21}{20} \; \zeta_{5}\right) \; \varepsilon^{-1} $ 
\\ \hline 
22 &
$e112|34|345|e5|55||$ &
$ 1/2 $ &
$ r_{1} \; r_{10} $ &
$ -\frac{1}{180} \; \varepsilon^{-5}+\frac{29}{720} \; \varepsilon^{-4}-\frac{217}{1440} \; \varepsilon^{-3}+\left(\frac{1019}{2880} - \frac{7}{30} \; \zeta_{3}\right) \; \varepsilon^{-2}+\left(-\frac{1903}{1920} + \frac{1}{40} \; \zeta_{4} + \frac{3}{5} \; \zeta_{3}\right) \; \varepsilon^{-1} $ 
\\ \hline 
23 &
$e112|34|355|45|e5||$ &
$ 1/2 $ &
$ r_{1} \; r_{10} $ &
$ -\frac{1}{180} \; \varepsilon^{-5}+\frac{29}{720} \; \varepsilon^{-4}-\frac{217}{1440} \; \varepsilon^{-3}+\left(\frac{1019}{2880} - \frac{1}{15} \; \zeta_{3}\right) \; \varepsilon^{-2}+\left(-\frac{1903}{1920} - \frac{1}{10} \; \zeta_{4} + \frac{3}{5} \; \zeta_{3}\right) \; \varepsilon^{-1} $ 
\\ \hline 
24 &
$e112|34|355|e4|55||$ &
$ 1/4 $ &
$ r_{1} \; r_{13} $ &
$ -\frac{1}{90} \; \varepsilon^{-5}+\frac{1}{24} \; \varepsilon^{-4}-\frac{13}{720} \; \varepsilon^{-3}-\frac{11}{480} \; \varepsilon^{-2}+\left(-\frac{1769}{2880} + \frac{7}{20} \; \zeta_{3}\right) \; \varepsilon^{-1} $ 
\\ \hline 
25 &
$e112|34|e33|5|555||$ &
$ 1/24 $ &
$ r_{1}^{2} \; r_{4} $ &
$ -\frac{13}{720} \; \varepsilon^{-4}+\frac{197}{2880} \; \varepsilon^{-3}-\frac{1}{48} \; \varepsilon^{-2}+\left(-\frac{223}{2304} - \frac{1}{48} \; \zeta_{3}\right) \; \varepsilon^{-1} $ 
\\ \hline 
26 &
$e112|34|e34|55|55||$ &
$ 1/8 $ &
$ r_{1} \; r_{14} $ &
$ -\frac{1}{72} \; \varepsilon^{-5}+\frac{11}{240} \; \varepsilon^{-4}+\frac{53}{1440} \; \varepsilon^{-3}+\left(-\frac{61}{192} + \frac{17}{60} \; \zeta_{3}\right) \; \varepsilon^{-2}+\left(\frac{157}{5760} + \frac{1}{20} \; \zeta_{4} - \frac{7}{40} \; \zeta_{3}\right) \; \varepsilon^{-1} $ 
\\ \hline 
27 &
$e112|34|e35|45|55||$ &
$ 1/2 $ &
$ r_{1} \; r_{13} $ &
$ -\frac{1}{144} \; \varepsilon^{-5}+\frac{71}{1440} \; \varepsilon^{-4}-\frac{101}{576} \; \varepsilon^{-3}+\left(\frac{1319}{5760} - \frac{11}{120} \; \zeta_{3}\right) \; \varepsilon^{-2}+\left(\frac{29}{3840} + \frac{1}{20} \; \zeta_{4} + \frac{11}{240} \; \zeta_{3}\right) \; \varepsilon^{-1} $ 
\\ \hline 
28 &
$e112|34|e55|445|5||$ &
$ 1/16 $ &
$ r_{1} \; r_{9} $ &
$ -\frac{1}{36} \; \varepsilon^{-5}+\frac{31}{360} \; \varepsilon^{-4}-\frac{13}{240} \; \varepsilon^{-3}+\left(-\frac{11}{288} - \frac{1}{60} \; \zeta_{3}\right) \; \varepsilon^{-2}+\left(-\frac{511}{2880} - \frac{1}{40} \; \zeta_{4} + \frac{23}{120} \; \zeta_{3}\right) \; \varepsilon^{-1} $ 
\\ \hline 
29 &
$e112|e3|334|5|555||$ &
$ 1/24 $ &
$ r_{1}^{3} $ &
$ -\frac{1}{384} \; \varepsilon^{-3}+\frac{5}{128} \; \varepsilon^{-2}-\frac{7}{32} \; \varepsilon^{-1} $ 
\\ \hline 
30 &
$e112|e3|344|55|55||$ &
$ 1/16 $ &
$ r_{1}^{2} \; r_{4} $ &
$ -\frac{1}{160} \; \varepsilon^{-4}+\frac{3}{80} \; \varepsilon^{-3}-\frac{53}{640} \; \varepsilon^{-2}+\left(\frac{59}{1280} + \frac{7}{80} \; \zeta_{3}\right) \; \varepsilon^{-1} $ 
\\ \hline 
31 &
$e112|e3|345|45|55||$ &
$ 1/8 $ &
$ r_{1}^{2} \; r_{3} $ &
$ -\frac{1}{480} \; \varepsilon^{-4}+\frac{11}{480} \; \varepsilon^{-3}-\frac{71}{640} \; \varepsilon^{-2}+\left(\frac{293}{1280} + \frac{7}{40} \; \zeta_{3}\right) \; \varepsilon^{-1} $ 
\\ \hline 
32 &
$e112|e3|444|555|5||$ &
$ 1/72 $ &
$ r_{1}^{3} $ &
$ -\frac{1}{192} \; \varepsilon^{-3}+\frac{5}{192} \; \varepsilon^{-2}-\frac{11}{384} \; \varepsilon^{-1} $ 
\\ \hline 
33 &
$e112|e3|445|455|5||$ &
$ 1/8 $ &
$ r_{1}^{2} \; r_{3} $ &
$ -\frac{1}{240} \; \varepsilon^{-4}+\frac{17}{480} \; \varepsilon^{-3}-\frac{173}{960} \; \varepsilon^{-2}+\left(\frac{1249}{1920} - \frac{3}{20} \; \zeta_{3}\right) \; \varepsilon^{-1} $ 
\\ \hline 
34 &
$e123|224|4|555|e5||$ &
$ 1/24 $ &
$ r_{1}^{2} \; r_{3} $ &
$ -\frac{1}{120} \; \varepsilon^{-4}+\frac{11}{320} \; \varepsilon^{-3}-\frac{3}{80} \; \varepsilon^{-2}+\left(\frac{401}{3840} - \frac{7}{40} \; \zeta_{3}\right) \; \varepsilon^{-1} $ 
\\ \hline 
35 &
$e123|224|5|445|5|e|$ &
$ 1/4 $ &
$ r_{1} \; r_{10} $ &
$ -\frac{1}{120} \; \varepsilon^{-5}+\frac{11}{240} \; \varepsilon^{-4}-\frac{49}{480} \; \varepsilon^{-3}+\left(\frac{47}{960} - \frac{1}{10} \; \zeta_{3}\right) \; \varepsilon^{-2}+\left(\frac{261}{640} - \frac{3}{20} \; \zeta_{4} + \frac{7}{60} \; \zeta_{3}\right) \; \varepsilon^{-1} $ 
\\ \hline 
36 &
$e123|234|45|45|5|e|$ &
$ 1/2 $ &
$ r_{1} \; r_{8} $ &
$ \frac{5}{3} \; \zeta_{5} \; \varepsilon^{-2}+\left(-\frac{25}{12} \; \zeta_{6} + \frac{1}{6} \; \zeta_{3}^{2}\right) \; \varepsilon^{-1} $ 
\\ \hline 
37 &
$e123|234|45|55|e5||$ &
$ 1/2 $ &
$ r_{1} \; r_{2} \; r_{3} $ &
$ -\frac{1}{10} \; \zeta_{3} \; \varepsilon^{-3}+\left(-\frac{3}{20} \; \zeta_{4} + \frac{11}{20} \; \zeta_{3}\right) \; \varepsilon^{-2}+\left(-\frac{3}{10} \; \zeta_{4} - \frac{7}{40} \; \zeta_{3} - \frac{1}{30} \; \zeta_{5}\right) \; \varepsilon^{-1} $ 
\\ \hline 
38 &
$e123|245|45|445||e|$ &
$ 1/4 $ &
$ r_{1} \; r_{2} \; r_{3} $ &
$ -\frac{1}{10} \; \zeta_{3} \; \varepsilon^{-3}+\left(-\frac{3}{20} \; \zeta_{4} + \frac{11}{20} \; \zeta_{3}\right) \; \varepsilon^{-2}+\left(-\frac{3}{10} \; \zeta_{4} - \frac{7}{40} \; \zeta_{3} + \frac{23}{60} \; \zeta_{5}\right) \; \varepsilon^{-1} $ 
\\ \hline 
39 &
$e123|e23|34|5|555||$ &
$ 1/12 $ &
$ r_{1}^{2} \; r_{3} $ &
$ -\frac{1}{288} \; \varepsilon^{-4}+\frac{25}{576} \; \varepsilon^{-3}-\frac{91}{384} \; \varepsilon^{-2}+\left(\frac{583}{1152} + \frac{1}{24} \; \zeta_{3}\right) \; \varepsilon^{-1} $ 
\\ \hline 
40 &
$e123|e23|44|55|55||$ &
$ 1/16 $ &
$ r_{1} \; r_{9} $ &
$ -\frac{1}{120} \; \varepsilon^{-5}+\frac{7}{240} \; \varepsilon^{-4}+\frac{11}{480} \; \varepsilon^{-3}+\left(-\frac{197}{960} + \frac{2}{15} \; \zeta_{3}\right) \; \varepsilon^{-2}+\left(\frac{443}{1920} + \frac{23}{40} \; \zeta_{4} - \frac{43}{60} \; \zeta_{3}\right) \; \varepsilon^{-1} $ 
\\ \hline 
41 &
$e123|e23|45|45|55||$ &
$ 1/8 $ &
$ r_{1} \; r_{11} $ &
$ -\frac{1}{360} \; \varepsilon^{-5}+\frac{17}{720} \; \varepsilon^{-4}-\frac{11}{160} \; \varepsilon^{-3}+\left(-\frac{587}{2880} + \frac{13}{30} \; \zeta_{3}\right) \; \varepsilon^{-2}+\left(\frac{10453}{5760} - \frac{1}{10} \; \zeta_{4} - \frac{101}{60} \; \zeta_{3}\right) \; \varepsilon^{-1} $ 
\\ \hline 
42 &
$e123|e24|33|5|555||$ &
$ 1/6 $ &
$ r_{1}^{2} \; r_{3} $ &
$ -\frac{7}{960} \; \varepsilon^{-4}+\frac{1}{30} \; \varepsilon^{-3}-\frac{11}{768} \; \varepsilon^{-2}+\left(-\frac{73}{512} + \frac{1}{24} \; \zeta_{3}\right) \; \varepsilon^{-1} $ 
\\ \hline 
43 &
$e123|e24|34|55|55||$ &
$ 1/4 $ &
$ r_{1} \; r_{13} $ &
$ -\frac{1}{360} \; \varepsilon^{-5}+\frac{1}{48} \; \varepsilon^{-4}-\frac{77}{1440} \; \varepsilon^{-3}+\left(-\frac{31}{960} + \frac{2}{15} \; \zeta_{3}\right) \; \varepsilon^{-2}+\left(\frac{2243}{5760} - \frac{7}{40} \; \zeta_{4} - \frac{1}{4} \; \zeta_{3}\right) \; \varepsilon^{-1} $ 
\\ \hline 
44 &
$e123|e24|35|45|55||$ &
$ 1 $ &
$ r_{1} \; r_{10} $ &
$ -\frac{1}{720} \; \varepsilon^{-5}+\frac{5}{288} \; \varepsilon^{-4}-\frac{347}{2880} \; \varepsilon^{-3}+\left(\frac{3037}{5760} - \frac{11}{60} \; \zeta_{3}\right) \; \varepsilon^{-2}+\left(-\frac{1323}{1280} + \frac{1}{10} \; \zeta_{4} + \frac{13}{24} \; \zeta_{3}\right) \; \varepsilon^{-1} $ 
\\ \hline 
45 &
$e123|e24|55|445|5||$ &
$ 1/4 $ &
$ r_{1} \; r_{11} $ &
$ -\frac{1}{240} \; \varepsilon^{-5}+\frac{13}{480} \; \varepsilon^{-4}-\frac{11}{192} \; \varepsilon^{-3}+\left(-\frac{239}{1920} + \frac{1}{12} \; \zeta_{3}\right) \; \varepsilon^{-2}+\left(\frac{1211}{1280} + \frac{1}{8} \; \zeta_{4} - \frac{97}{120} \; \zeta_{3}\right) \; \varepsilon^{-1} $ 
\\ \hline 
46 &
$e123|e45|334|5|55||$ &
$ 1/8 $ &
$ r_{1} \; r_{11} $ &
$ -\frac{1}{90} \; \varepsilon^{-5}+\frac{1}{20} \; \varepsilon^{-4}-\frac{17}{360} \; \varepsilon^{-3}+\left(-\frac{19}{240} + \frac{1}{60} \; \zeta_{3}\right) \; \varepsilon^{-2}+\left(-\frac{49}{1440} + \frac{1}{40} \; \zeta_{4} - \frac{3}{40} \; \zeta_{3}\right) \; \varepsilon^{-1} $ 
\\ \hline 
47 &
$e123|e45|344|55|5||$ &
$ 1/8 $ &
$ r_{1} \; r_{12} $ &
$ -\frac{1}{360} \; \varepsilon^{-5}+\frac{1}{48} \; \varepsilon^{-4}-\frac{77}{1440} \; \varepsilon^{-3}+\left(-\frac{31}{960} + \frac{2}{15} \; \zeta_{3}\right) \; \varepsilon^{-2}+\left(\frac{2243}{5760} - \frac{7}{40} \; \zeta_{4} - \frac{1}{4} \; \zeta_{3}\right) \; \varepsilon^{-1} $ 
\\ \hline 
48 &
$e123|e45|345|45|5||$ &
$ 1/4 $ &
$ r_{1} \; r_{2} \; r_{3} $ &
$ -\frac{1}{6} \; \zeta_{3} \; \varepsilon^{-3}+\left(\frac{1}{4} \; \zeta_{4} + \frac{7}{12} \; \zeta_{3}\right) \; \varepsilon^{-2}+\left(-\frac{1}{2} \; \zeta_{4} - \frac{5}{8} \; \zeta_{3} + \frac{2}{3} \; \zeta_{5}\right) \; \varepsilon^{-1} $ 
\\ \hline 
49 &
$e123|e45|444|555|||$ &
$ 1/72 $ &
$ r_{1}^{3} $ &
$ -\frac{1}{192} \; \varepsilon^{-3}+\frac{5}{192} \; \varepsilon^{-2}-\frac{11}{384} \; \varepsilon^{-1} $ 
\\ \hline 
50 &
$e123|e45|445|455|||$ &
$ 1/8 $ &
$ r_{1} \; r_{10} $ &
$ -\frac{1}{360} \; \varepsilon^{-5}+\frac{1}{48} \; \varepsilon^{-4}-\frac{77}{1440} \; \varepsilon^{-3}+\left(-\frac{31}{960} - \frac{1}{30} \; \zeta_{3}\right) \; \varepsilon^{-2}+\left(\frac{2243}{5760} - \frac{1}{20} \; \zeta_{4} - \frac{1}{4} \; \zeta_{3}\right) \; \varepsilon^{-1} $ 
\\ \hline 
\hline
\end{longtable}

\def\arraystretch{1.5}
\setlength{\tabcolsep}{5pt}
\begin{longtable}{| c | c || c | c | }
\caption{Values of the factors $r_i(n)$ in 
\ref{tableG2}} \label{tabler}\\
\hline
$i$ & $\qquad \qquad \qquad r_i(n)\qquad \qquad \qquad $ & $i$ & $r_i(n)$\\
\hline\hline
    ${1}$ & $ (n+2)/3  $ & ${9}$ & $ (3 n^3+24 n^2+80 n+136)/243  $\\ \hline
    ${2}$ & $ (n+8)/9  $ & ${10}$ & $ (7 n^2+72 n+164)/243$\\ \hline
    ${3}$ & $ (5 n+22)/27  $ & ${11}$ & $ (11 n^2+76 n+156)/243  $ \\ \hline
    ${4}$ & $ (n^2+6 n+20)/27  $ & ${12}$ & $ (n^3+10 n^2+72 n+160)/243  $ \\ \hline
    ${5}$ & $ (3 n^2+22 n+56)/81  $ & ${13}$ & $ (n^3+14 n^2+76 n+152)/243  $ \\ \hline
    ${6}$ & $ (n^2+20 n+60)/81  $ & ${14}$ & $ (n^3+18 n^2+80 n+144)/243  $ \\ \hline
    ${7}$ & $ (n^3+8 n^2+24 n+48)/81  $ & ${15}$ & $ (n^4+10 n^3+40 n^2+80 n+112)/243  $ \\ \hline

    ${8}$ & $ (2 n^2+55 n+186)/243  $ & &\\ \hline
\hline
\end{longtable}

\section{Extended  {}'t Hooft condition for separate diagrams \label{app2}}

In this Appendix we discuss an extension\footnote{We do not claim that the
extension is an original contribution of us. In fact, at least for IR-finite
diagrams it is well-known among experts since long. For instance, very
recently similar  considerations have been effectively employed in \cite{Bork:2015zaa} to
study  divergences in maximal supersymmetric Yang-Mills theories in diverse dimensions.
 }    of the well-known {}'t Hooft
constraints  originally suggested in \cite{'tHooft:1973mm} for   global
renormalization constants (that is ones including all contributions up to 
some number of loops) to  a  case when one deals with a separate Feynman integral.

Let $\Gamma$ be a particular L-loop OPI Feynman diagram without any IR
(sub)divergences\footnote{This constraint will be relaxed later.}.  Without essential
loss of generality we assume that $\langle\Gamma\rangle(Q^2,\mu^2)$ is a
scalar integral depending on the external momentum $Q$ via its square, $Q^2 =
Q_\nu Q^\nu$.  In addition, we introduce the renormalization scale parameter
$\mu$ into the definition of every bare dimensionally regulated FI by providing
it with a factor $(\mu^2)^{L\, \ep}$. 

The renormalized version of the corresponding  Feynman 
integral can be generically  written as
\beq
R \,\langle\Gamma\rangle(Q^2,\mu^2) = \langle\Gamma\rangle(Q^2,\mu^2) + Z_\Gamma +
\fbox{$
\sum_{\gamma}Z_\gamma \langle\Gamma/\gamma\rangle(Q^2)
{} + \dots$}
\label{R}
\eeq
Here $Z_\gamma$ is the  UV Z-factor corresponding to a OPI subgraph
$\gamma$ of $\Gamma$, $Z_\Gamma$ is the UV counterterm for the very 
FI  $\langle\Gamma\rangle$   and dots stand for contributions with two and
more UV subtractions.   

Every  particular term in the  boxed part of 
eq.~(\ref{R}) is a product of some Z-factors and a reduced FI, the latter
by construction includes a factor  $(\mu^2)^{n \ep}$, with $n$ being its loop
number.  

The finiteness of the left part of eq. (\ref{R}) together with the fact that
the $Z_\gamma$ has no dependence on $\mu$ leads to an a number of interesting
consequences. For instance, if $L=2$ then only the knowledge of the pole parts
of the {\em one-loop} subgraphs of $\Gamma$ as well as one-loop reduced FI
$\langle \Gamma/\gamma\rangle$ allows one to construct the leading $1/\ep^2$
poles of the FI $\langle \Gamma \rangle$ and the counterterm
$Z_\Gamma$. By induction, one could easily infer that for arbitrary number of
loops $L$ the leading $1/\ep^L$ poles of both the FI 
$\langle\Gamma\rangle$ and the corresponding counterterm $Z_\Gamma$ can be
completely restored from the pole parts (read UV counterterms) of properly
constructed set of one-loop FIs. The set includes all graphs of the form
$\gamma/\gamma'$, with $\gamma$ and $\gamma'$ being two OPI  subgraphs of
$\Gamma$ such that $\gamma' \subset \gamma$ and $L_\gamma -L_{\gamma'} =1$.

In the same way one could infer {\em subleading} poles of order $1/\ep^{L-1}$ 
exclusively from knowledge of Z-factors from similarly constructed set of 
two-loop FIs. And so on and forth. This is, obviously, the diagram-wise
formulation of the {}'t Hoof constraints. 

Another  simple (but still   useful)  observation 
is  that the knowledge of $Z_\Gamma$ and all the boxed terms in the r.h.s. of  (\ref{R})
is enough to {\em completely}  restore the pole part of the original bare FI  
$\langle\Gamma\rangle$.  

In fact, all the above considerations are easily  generalized
for a case when FI $\langle\Gamma\rangle$ is suffering from IR divergences
in addition to  UV ones\footnote{This statement is only valid for Euclidean case, as
the very $R^*$-operation is not suitable to deal with more complicated
(collinear, etc.)  IR singularities which might appear in  Minkowskian FIs.}. Indeed, 
as it should be clear from the general discussion of section \ref{sec:calcgen} it
suffices to employ the $R^*$-operation instead of the usual $R$-one. 

Finally, let us  now assume that the FI $ \langle\Gamma\rangle $ is a massless
five-loop propagator-like FI.  Combining two facts: (i) 5-loop Z-factors are all
computable in terms of 4-loop p-integrals and (ii) every reduced FI in the r.h.s of
(\ref{R}) is a p-integral with its loop number not exceeding 4, we arrive to a
conclusion that the  pole part of $ \langle\Gamma\rangle $ is expressible in terms of
4-loop p-integrals. 

As an example we present here complete expression for pole part of the five loop p-integral from section~\ref{subsec:diag1}:
\begin{equation}
K\left(\begin{matrix}
\includegraphics{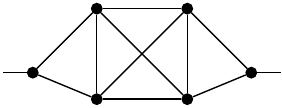}
\end{matrix}\right)
\end{equation}
Taking into account that 
\begin{equation}
KR'\left(\begin{matrix}
\includegraphics{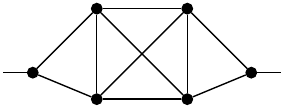}
\end{matrix}\right)=KR^{*\prime} \left(\begin{matrix}
\includegraphics{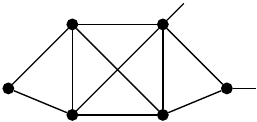} 
\end{matrix}\right)\,,
\label{diag1Rstar}
\end{equation}
and r.h.s. is computable in terms of 4-loop p-integrals, and  expanding $R'$ operation in the l.h.s  of the \eqref{diag1Rstar} we arrive to the following relation:
\begin{equation}
\begin{aligned}
K\left(\begin{matrix}
\includegraphics{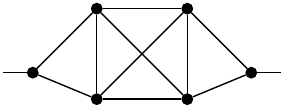}
\end{matrix}\right)=&K\left(R^{*\prime} \left(\begin{matrix}
\includegraphics{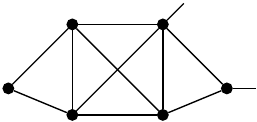}
\end{matrix}\right)
+KR'\left(\begin{matrix}
\includegraphics{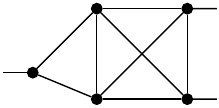}
\end{matrix} \right) 
\begin{matrix}
\includegraphics{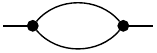}
\end{matrix} \,+\right.\\
&\qquad\left.+KR'\left(
\begin{matrix}
\includegraphics{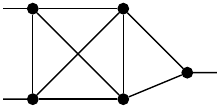}
\end{matrix}\right)\begin{matrix}
\includegraphics{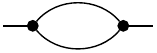}
\end{matrix}
+ KR'\left(
\begin{matrix}
\includegraphics{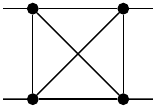}
\end{matrix}
\right)
\includegraphics{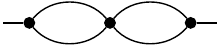}
\right)
\end{aligned}
\label{diag1Rstar2}
\end{equation}
Here all terms of r.h.s of~\eqref{diag1Rstar2} can be expressed in terms of 4-loop p-integrals.

\ice{
Let us now

means that the pole part in $\ep = (4-D)/2$ of $\langle\Gamma\rangle(Q^2)$ is
completely fixed by poles in $\ep$ which appear in UV subtractions (the
boxed terms in (\ref{R})). On the other hand, the UV subtractions
could, obviously, contain $L'$-loop Z-factors with $L' \le L+1$ and
the reduced p-integrals like $\langle \Gamma/\gamma\rangle(Q^2)$ with the loop
number {\em not exceeding} L! Applying Theorem 2 we arrive  at the 
conclusion that the pole part of $\langle\Gamma\rangle(Q^2)$ (and, consequently,
its absorptive part) is completely expressed via L-loop p-integrals
{\em only}.
}

\section{$1/n$-expansion }
\label{error_in_29}

In paper \cite{oneovern3} conformal bootstrap technique was applied to calculate $1/n$-expansion of the critical exponent $\eta$ up to $1/n^3$ term:
\begin{equation}
\eta = \frac{\eta_1}{n} + \frac{\eta_2}{n^2}+ \frac{\eta_3}{n^3}+{\cal O}\left(\frac{1}{n^4}\right)\,.
\label{etan}
\end{equation}
It is possible to compare results obtained using $\varepsilon$-expansion with results of $1/n$-expansion for this exponent: while $\varepsilon$-expansion is an exact function of $n$, $1/n$-expansion calculated in \cite{oneovern3} is an exact function of $\varepsilon$. Thus twofold series of both expansions must coincide.

\begin{figure}[h]
\centering
\includegraphics{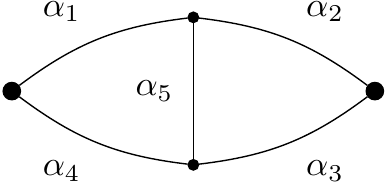}
\caption{T-bubble graph contributing to $\eta_3$  ($\alpha_1=\alpha_4=1$, $\alpha_2=\alpha_3=d/2-1$, $\alpha_5=d/2-1+\Delta$)}
\label{fig:tbuble1n}
\end{figure}

Unfortunately, $\eta_3$ in \cite{oneovern3} contain misprint, so we present corrected version here:
\begin{equation}
\eta_1 = -\frac{4 \Gamma(d-2)}{\Gamma(2-d/2)\Gamma(d/2-2)\Gamma(d/2-1)\Gamma(d/2+1)},
\label{etan1}
\end{equation}
\begin{equation}
\frac{\eta_2}{\eta_1^2}=\frac{d^2-3d+4}{4-d} R_0 +\frac{1}{d}+\frac{1}{d-2}+\frac{9}{4-d}+\frac{4}{(4-d)^2}-2-d
\label{etan2}
{},
\end{equation}
where $R_0=\psi(d-2)+\psi(2-d/2)-\psi(2)-\psi(d/2-2)$ and $\psi(x)=\frac{d}{dx} \ln \; \Gamma(x)$.  Furthermore, 
\begin{equation}
\begin{aligned}
\frac{\eta_3}{\eta_1^3} = & \frac{3 d^2(d-2)(2d-5) I(d/2)S_3}{4(4-d)^2}+\frac{2}{3}\frac{d^2(d-2)(d-3)^2(3S_0S_1-S_0^3-S_2)}{(4-d)^3} +\\
&+35+\frac{13}{2}d+d^2-\frac{177}{4-d}+\frac{134}{(4-d)^2}+\frac{232}{(4-d)^2}-\frac{128}{(4-d)^3}+\frac{9}{d-2}+\frac{2}{(d-2)^2}+\frac{2}{d^2}+\\
&+\frac{B}{2}\left(66+7d+d^2-\frac{374}{4-d}+\frac{408}{(4-d)^2}+\frac{128}{(4-d)^3}+\frac{4}{d-2}+\frac{6}{d}\right)+\frac{B^2}{2}\left(20-\frac{100}{4-d}+\frac{128}{(4-d)^2}\right)+\\
&+\frac{S_3}{2}\left(-45-5d+\frac{7}{4}d^2+\frac{254}{4-d}-\frac{256}{(4-d)^2}-\frac{384}{(4-d)^3}+\frac{512}{(4-d)^4}\right)+\frac{S_4}{2}\left(14+4d+2d^2-\frac{60}{4-d}\right)+\\
&+\frac{BS_3}{2}\left(-45-\frac{13}{2}d-\frac{1}{2}d^2+\frac{272}{4-d}-\frac{432}{(4-d)^2}+\frac{256}{(4-d)^3}\right) 
{},
\label{etan3}
\end{aligned}
\end{equation}
where 
\begin{equation}
\begin{aligned}
B \,&=\psi(2-d/2)+\psi(d-2)-1+\gamma_E-\psi(d/2-2),\\
S_0 &=\psi(2-d/2)+\psi(d-2)+\gamma_E-\psi(d/2-1),\\
S_1 &=\psi'(2-d/2)-\psi'(d-2)-\zeta(2)+\psi'(d/2-1),\\
S_2 &=\psi''(2-d/2)+\psi''(d-2)+2\zeta(3)-\psi''(d/2-1),\\
S_3 &=\psi'(d/2-1)-\psi'(1),\\
S_4 &=\psi''(2-d/2)-\psi'(d-2),
\label{etansubs}
\end{aligned}
\end{equation}
$\gamma_E$ is Euler constant and value $I(d)$ is determined from the relation:
\begin{equation}
\Pi(d,\Delta) =  \Pi(d,0)\left(1+I(d) \, \Delta + {\cal O}(\Delta^2)\right)
{}.
\end{equation}
Here $\Pi(d,\Delta)$ is value of the diagram on Fig.\ref{fig:tbuble1n} (in x-space)  with $\alpha_1=\alpha_4=1$, $\alpha_2=\alpha_3=d/2-1$, $\alpha_5=d/2-1+\Delta$
 
Given  the value of the $I(d)$ for any $d$ one can construct $1/n$
expansion for arbitrary space dimension. 
\ice{
unfortunately up to now this diagram
didn't calculated for arbitrary $d$, only some particular cases are known. 
For example, for $d=3$ exact value $I(3)=3\psi''(1/2)/2\pi^2+2 \ln 2$ calculated
in \cite{oneovern3}, while for $d=2+2\varepsilon$ \cite{oneovern3,Imu2plusE}
and $d=4-2\varepsilon$ \cite{uniquenessK,Kazakov:1985tmf} only
$\varepsilon$-expansion is known.  
}
The value of $I(4-2\varepsilon)$ can be
extracted from \cite{Kazakov:1985tmf} with the result:
\begin{equation}
I(4-2\varepsilon)=-\frac{5 \zeta(5)}{2 \zeta(3)}\varepsilon + \left(\frac{15\zeta(4)\zeta(5)-25\zeta(3)\zeta(6)+10\zeta(3)^3}{4 \zeta(3)^2}\right)\varepsilon^2+{\cal O}(\varepsilon^3)
{}.
\label{Imu}
\end{equation}

Combining \eqref{etan}--\eqref{etansubs} with \eqref{Imu} and expanding it in
$\varepsilon$ up to $\varepsilon^6$ term, and, from another hand, expanding
\eqref{etae6} in $1/n$ up to $1/n^3$ term we arrive at two {\em identical}
expansions with 
\begin{equation}
\begin{split}
\eta_1&=2 \; \varepsilon^{2}- \varepsilon^{3}-\frac{5}{2} \; \varepsilon^{4}+\left(-\frac{13}{4} + 4 \; \zeta_{3}\right) \; \varepsilon^{5}+\left(-\frac{29}{8} - 2 \; \zeta_{3} + 6 \; \zeta_{4}\right) \; \varepsilon^{6}+{\cal O}(\varepsilon^7)\\
\eta_2&=-28 \; \varepsilon^{2}+86 \; \varepsilon^{3}+\left(-35 - 176 \; \zeta_{3}\right) \; \varepsilon^{5}+\left(-\frac{243}{4} + 488 \; \zeta_{3} - 264 \; \zeta_{4}\right) \; \varepsilon^{6}+{\cal O}(\varepsilon^7)\\
\eta_3&=320 \; \varepsilon^{2}-1984 \; \varepsilon^{3}+\left(2732 - 960 \; \zeta_{3}\right) \; \varepsilon^{4}+\left(686 + 9440 \; \zeta_{3} - 1440 \; \zeta_{4} + 2560 \; \zeta_{5}\right) \; \varepsilon^{5}+\\
&\;\;+\left(799 - 28104 \; \zeta_{3} + 14160 \; \zeta_{4} + 1024 \; \zeta_{3}^{2} + 6400 \; \zeta_{6} - 24400 \; \zeta_{5}\right) \; \varepsilon^{6}+{\cal O}(\varepsilon^7)
\end{split}
\end{equation}
Two comments are required here. First, equality of the twofold series produces
three independent relation for six loop diagram values (only one six loop
diagram contributes to $1/n$ term, to $1/n^2$ contributes 20 diagrams, and to
$1/n^3$ -- 50 diagrams, i.e. all six loop diagrams). So comparison with the $1/n$
expansion should  be considered as a really strong check of our six loop results.
Second, actually only the first term from $I(d)$ (of order $\varepsilon$)
is required for six loops, the next term of $I(d)$ will contribute to seven
loop term, but to get the same kind of relations (which touch all seven loop
diagrams) one would need to calculate $1/n^4$ contribution to \eqref{etan}.
\ice{
 Anyway,
even partial check of the future seven loop term is very important.
}

\end{document}